**Comments on "$\eta \to \gamma\gamma$ Decay Width via the Primakoff Cross Section"**

In a recent article by Rodrigues et al. [1], $\eta$ photo-nuclear data taken by Browman et al. [2] in 1974 were refit as a means to extract the $\eta \to \gamma\gamma$ decay width. The key assertion in the author's analysis is attributing backgrounds in the $\eta$ angular distribution (see Fig. 2 in [2]) to nuclear incoherent production only. The Cornell angular distributions were fitted using Primakoff and nuclear coherent distributions taken from the original analysis [2], and a nuclear incoherent background calculated using a nuclear cascade model [1]. Rodrigues et al. obtain $\Gamma(\eta \to \gamma\gamma) = 0.476(62)$ keV, leading them to conclude that the long-standing discrepancy between the Cornell value of 0.324(46) keV and the average of collider experiments 0.510(26) keV [3] has been resolved. We disagree with this assertion, and question the validity of the fitting procedure utilized by Rodrigues, et al. In their analysis, backgrounds in the Cornell angular distribution are attributed solely to nuclear incoherent production, and the other important backgrounds have been neglected. A reanalysis of the Cornell Primakoff data requires a detailed understanding of the hadronic and accidental backgrounds in the data, and the authors of [1] have not accomplished this task.

The Cornell experiment was performed more than thirty years ago with an untagged bremsstrahlung photon beam. Two conventional lead glass hodoscopes were used to detect decay photons. Due to the unknown event timing and energy of the incident photon, and the relatively poor position and energy resolution of the lead glass hodoscopes, there was significant background in the $\eta$ angular distributions (see Figs. 1 and 2 in [2]). Furthermore, as stated in [2] and [4], this background varies with target, and is different for the $\pi^0$ and $\eta$ experiments (see Ref. 8 in [2]). We can surmise that backgrounds in the Cornell experiment result from accidentals and beam related backgrounds, hadronic backgrounds, as well as nuclear incoherent production. Due to the complexity of the background, Browman et al. assumed an isotropic angular distribution for the background, and did not use a model for nuclear incoherent scattering with Pauli-blocking which was available at the time [5]. Browman et al. later measured $\Gamma(\pi^0 \to \gamma\gamma) = 7.92 \pm 0.42$ eV [5] using the same apparatus, and their result is in good agreement with the world average of $7.7 \pm 0.6$ eV [3]. In this later experiment, the distance between the target and the lead glass hodoscopes was increased by factor of two, which significantly reduced backgrounds in the $\pi^0$ experiment relative to the $\eta$ experiment (see Ref. 8 in [2]), as well as improving the $\pi^0$ angular resolution.

Evidence that the fits by Rodrigues *et al.* are inconsistent can be seen in table I of Ref. 1, where the fitted nuclear coherent amplitudes and interference angles between Primakoff and nuclear coherent amplitudes are listed for the Be and Cu targets. The target dependence of these parameters is expected to be relatively weak, and for the Cornell $\pi^0$ and $\eta$ analysis the same parameters were applied for different targets. For the high precision pion Primakoff experiment at Jefferson lab, PRIMEX, we observe a modest difference between $^{12}$C and $^{208}$Pb for these parameters. Preliminary PRIMEX results for the fitted nuclear coherent amplitudes for carbon and lead are $1.78 \pm .04$, and $1.2 \pm .1$, respectively. The fitted interference angles for carbon and lead are $.98 \pm .05$ rad and $1.14 \pm .08$, respectively. However, in [1] the ratio of fitted nuclear coherent amplitudes for Cu and Be targets is approximately seven, and the fitted interference angle is constructive for Be and destructive for Cu. There is no physical explanation for the strong target dependence of these parameters.

In summary, we conclude that the discrepancy between the collider experiments and the old Primakoff measurement on the $\eta$ decay width is unresolved. The background assumption employed by Rodrigues, *et al.* is simplistic; other important backgrounds have been neglected. The sensitivity of their result to the assumed background model remains unknown. While a reanalysis of the Cornell data is not impossible, the task is problematic given that the data were taken over 30 years ago and many experimental details have been lost. The only solution for solving this puzzle is to perform a new Primakoff experiment with a tagged photon facility and better calorimetry to control and minimize the background and the resolutions in the experiment.

L. Gan[1], A. Gasparian[2], S. Gevorkyan[3], and R. Miskimen[4]

1. University of North Carolina, Wilmington, NC 28403

2. North Carolina A&T State University, Greensboro, NC 27411

3. Joint Institute for Nuclear Research, Dubna, Russia

4. University of Massachusetts, Amherst, MA 01003